\documentclass[aps,prl,twocolumn,showpacs,floatfix,superscriptaddress]{revtex4}
\usepackage{graphicx}
\usepackage{hyperref}
\begin{document}
\flushbottom
\title{Superconductivity and magnetism on flux grown single crystals of NiBi$_3$}

\author{B. Silva}
\affiliation{Laboratorio de Bajas Temperaturas, Departamento de F\'isica de la Materia
Condensada, Instituto Nicol\'as Cabrera and
Condensed Matter Physics Center, Universidad Aut\'onoma de Madrid, E-28049 Madrid,
Spain}

\author{R.F. Luccas}
\affiliation{Laboratorio de Bajas Temperaturas, Departamento de F\'isica de la Materia
Condensada, Instituto Nicol\'as Cabrera and
Condensed Matter Physics Center, Universidad Aut\'onoma de Madrid, E-28049 Madrid,
Spain}

\author{N.M. Nemes}
\affiliation{Departamento de Fisica Aplicada III, GFMC, Universidad Complutense de Madrid, E-28040 Madrid, Spain}

\author{J. Hanko}
\affiliation{Laboratorio de Bajas Temperaturas, Departamento de F\'isica de la Materia
Condensada, Instituto Nicol\'as Cabrera and
Condensed Matter Physics Center, Universidad Aut\'onoma de Madrid, E-28049 Madrid,
Spain}

\author{M.R. Osorio}
\affiliation{Laboratorio de Bajas Temperaturas, Departamento de F\'isica de la Materia
Condensada, Instituto Nicol\'as Cabrera and
Condensed Matter Physics Center, Universidad Aut\'onoma de Madrid, E-28049 Madrid,
Spain}

\author{P. Kulkarni}
\affiliation{Laboratorio de Bajas Temperaturas, Departamento de F\'isica de la Materia
Condensada, Instituto Nicol\'as Cabrera and
Condensed Matter Physics Center, Universidad Aut\'onoma de Madrid, E-28049 Madrid,
Spain}

\author{F. Mompean}
\affiliation{Instituto de Ciencia de Materiales de Madrid, Consejo Superior de Investigaciones Cient\'{\i}ficas (ICMM-CSIC), Sor Juana In\'es de la Cruz 3, 28049 Madrid, Spain.}
\affiliation{Unidad Asociada de Bajas Temperaturas y Altos Campos Magn\'eticos, UAM/CSIC, Cantoblanco, E-28049 Madrid, Spain}

\author{M. Garc{\'i}a-Hern{\'a}ndez}
\affiliation{Instituto de Ciencia de Materiales de Madrid, Consejo Superior de Investigaciones Cient\'{\i}ficas (ICMM-CSIC), Sor Juana In\'es de la Cruz 3, 28049 Madrid, Spain.}
\affiliation{Unidad Asociada de Bajas Temperaturas y Altos Campos Magn\'eticos, UAM/CSIC, Cantoblanco, E-28049 Madrid, Spain}

\author{M.A. Ramos}
\affiliation{Laboratorio de Bajas Temperaturas, Departamento de F\'isica de la Materia
Condensada, Instituto Nicol\'as Cabrera and
Condensed Matter Physics Center, Universidad Aut\'onoma de Madrid, E-28049 Madrid,
Spain}
\affiliation{Unidad Asociada de Bajas Temperaturas y Altos Campos Magn\'eticos, UAM/CSIC, Cantoblanco, E-28049 Madrid, Spain}

\author{S. Vieira}
\affiliation{Laboratorio de Bajas Temperaturas, Departamento de F\'isica de la Materia
Condensada, Instituto Nicol\'as Cabrera and
Condensed Matter Physics Center, Universidad Aut\'onoma de Madrid, E-28049 Madrid,
Spain}
\affiliation{Unidad Asociada de Bajas Temperaturas y Altos Campos Magn\'eticos, UAM/CSIC, Cantoblanco, E-28049 Madrid, Spain}

\author{H. Suderow$^*$}
\affiliation{Laboratorio de Bajas Temperaturas, Departamento de F\'isica de la Materia
Condensada, Instituto Nicol\'as Cabrera and
Condensed Matter Physics Center, Universidad Aut\'onoma de Madrid, E-28049 Madrid,
Spain}
\affiliation{Unidad Asociada de Bajas Temperaturas y Altos Campos Magn\'eticos, UAM/CSIC, Cantoblanco, E-28049 Madrid, Spain}

\begin{abstract}
We present resistivity, magnetization and specific heat measurements on flux grown single crystals of NiBi$_3$. We find typical behavior of a type-II superconductor, with, however, a sizable magnetic signal in the superconducting phase. There is a hysteretic magnetization characteristic of a ferromagnetic compound. By following the magnetization as a function of temperature, we find a drop at temperatures corresponding to the Curie temperature of ferromagnetic amorphous Ni. Thus, we assign the magnetism in NiBi$_3$ crystals to amorphous Ni impurities.
\end{abstract}
 \date{\today} \maketitle

\section{Introduction}
The interplay between superconductivity and ferromagnetism has been one of the most fruitful areas of debate during recent years \cite{Canfield1998,Kivelson2001,Aoki2001,Coronado10}. The superconducting state is generally considered to be sensitive to small concentrations of magnetic impurities, which induce Cooper pair-breaking \cite{Matthias1953,Chervenak1995,Garret1998,Smith2000}. But in some cases, long-range magnetic order may coexist with superconductivity\cite{flouquet2002ferromagnetic,Bulaevski85}. This leads to rather unique effects arising from the interplay between the two phenomena. Successive transitions between superconductivity and other ordered states are observed in Chevrel phase and borocarbide compounds \cite{Fertig1977,Ott1978,Lynn1985,Prozorov2008,PhysRevLett.102.237002,PhysRevLett.96.027003,Galvis20121076,Crespo2006471}. Superconducting properties have been shown to be enhanced by an external field \cite{Maple1986,Sangiao11,Guillamon2010771,Suderow01,Cordoba13}. In heavy fermions and in hybrid proximity ferromagnetic-superconducting structures, magnetism can induce unconventional p-wave or other forms of complex superconductivity\cite{Buzdin05,Steglich1979, Brison2000, Steglich2005, Steglich2012}.

NiBi$_3$ is an intermetallic alloy known to be a type-II superconductor \cite{Fujimori2000} with a critical temperature of about $4$ K. Normally, Ni tends to lose its magnetic moment within this type of compounds\cite{Fujimori2000}. However, recent work shows coexistence of ferromagnetic-like signals with superconductivity in polycrystalline samples\cite{Pineiro2011}. Further work shows that ferromagnetism is absent in bulk single crystal samples below $300$ K, but that some kind of fluctuations do exist below $150$ K just at the surface \cite{Zhu2012}. In the same spirit, other authors remark the absence of ferromagnetic behavior in bulk single crystals, but show ferromagnetic and superconducting features in nanostructures. They highlight confinement effects which eventually modify the electronic band structure \cite{Herrmannsdorfer2011}. One of the main concerns when fabricating this kind of samples is that pure Ni inclusions can remain within the crystal, thus yielding a non-zero magnetic moment. 

In this work, we have studied the superconducting properties of high quality NiBi$_3$ single crystals grown by the flux growth method. We have carried out resistivity, magnetization and specific heat measurements, with the aim to understand the magnetic and superconducting behavior of this system. We indeed find a ferromagnetic signal, also in the superconducting phase, and discuss temperature dependence of magnetization, resistivity and specific heat.

\section{Experimental details}

The sample was grown in excess of Bi flux
\cite{Canfield1992,Canfield95,CanfieldBook}. 90\% high purity Bi (Alfa Aesar 99.99\%) and 10\% of Ni (Alfa Aesar 99.99\%) were introduced in an ampoule of quartz and sealed under inert gas atmosphere. Ampoules were heated for 4 hours until 1100$^o$C, maintained 100 hours at this temperature, and then cooled to 300$^o$C in 300 hours, where it remained for another 100 hours. Ampoules were then taken out of the furnace and rapidly centrifuged to remove the Bi flux. We obtained small needles of 0.1 mm$\times$0.1 mm$\times$1.5 mm as shown in Fig.\ 1. Most of these needles were joined together with residual Bi flux left over after centrifuging. We made powder X-ray diffraction on an arrangement of needles milled down to powder. We used an X'Pert PRO Theta/2Theta difractometer with primary monochromator and fast X'Celerator detector. We measured the resistivity making four contacts on a single needle. Specific heat was determined in PPMS system of Quantum Design. To ensure proper thermalization of the whole sample, a large amount of needles (6.2 mg) were crushed down into a pellet using a force of 5.5 tons during 6 minutes. We made magnetization hysteresis loops, M(H), at constant temperature below and above $T_c$ with a SQUID-magnetometer (MPMS) using single needles aligned parallel to the applied magnetic field.

\begin{figure}
\includegraphics[width=0.50\textwidth]{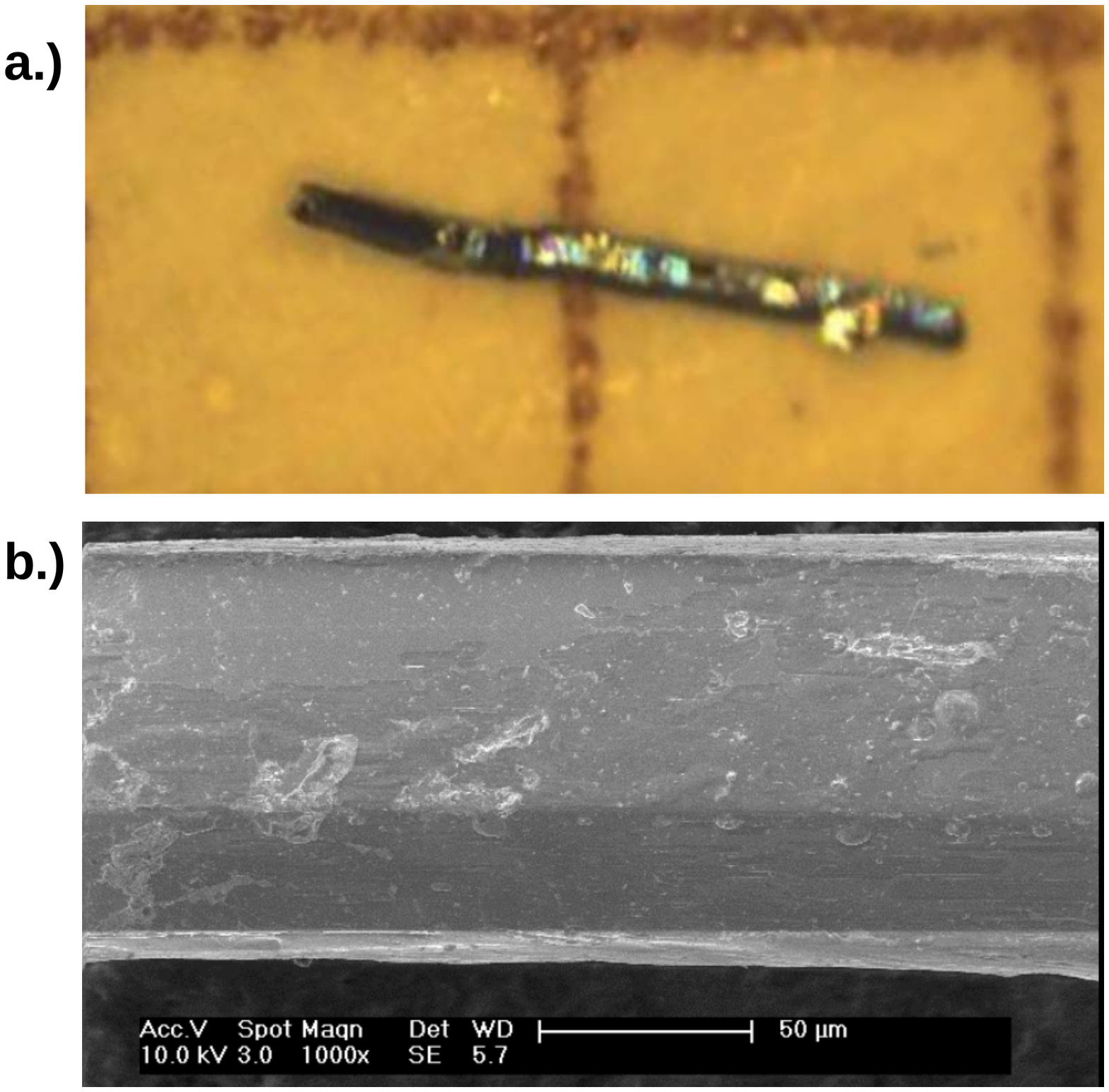}
\caption{a.) NiBi$_3$ needle of $\approx$ 1.2 mm length on a millimeter paper. The needle was separated mecanically from a bunch of NiBi$_3$ needles joined together by Bi flux. b.) Scanning Electron Microscope (SEM) picture of one needle with a crossection of 0.1 mm $\times$ 0.1 mm. Some Bi flux bubbles can be identified.}
\end{figure}

\section{Results and discussion}

NiBi$_3$ has an orthorombic CaLiSi$_2$-type crystal structure with space group pnma. X-ray powder diffractograms are shown in Fig.\ 2. We find single crystalline NiBi$_3$. The difffractogram also show a peak from Bi flux used during growth (bubbles in Fig.\ 1b). No indication for presence of residual Ni was found in X-ray scattering. Using Rietveld refinement (X'PERT HighScore Plus software), the unit cell parameters were found to be: a = 8.884 \AA , b = 4.097 {\AA} and c = 11.490 \AA. The inset in Fig.\ 2 shows the NiBi$_3$ crystal structure\cite{Ruck06}.

\begin{figure}
	\centering
		\includegraphics[width=0.50\textwidth]{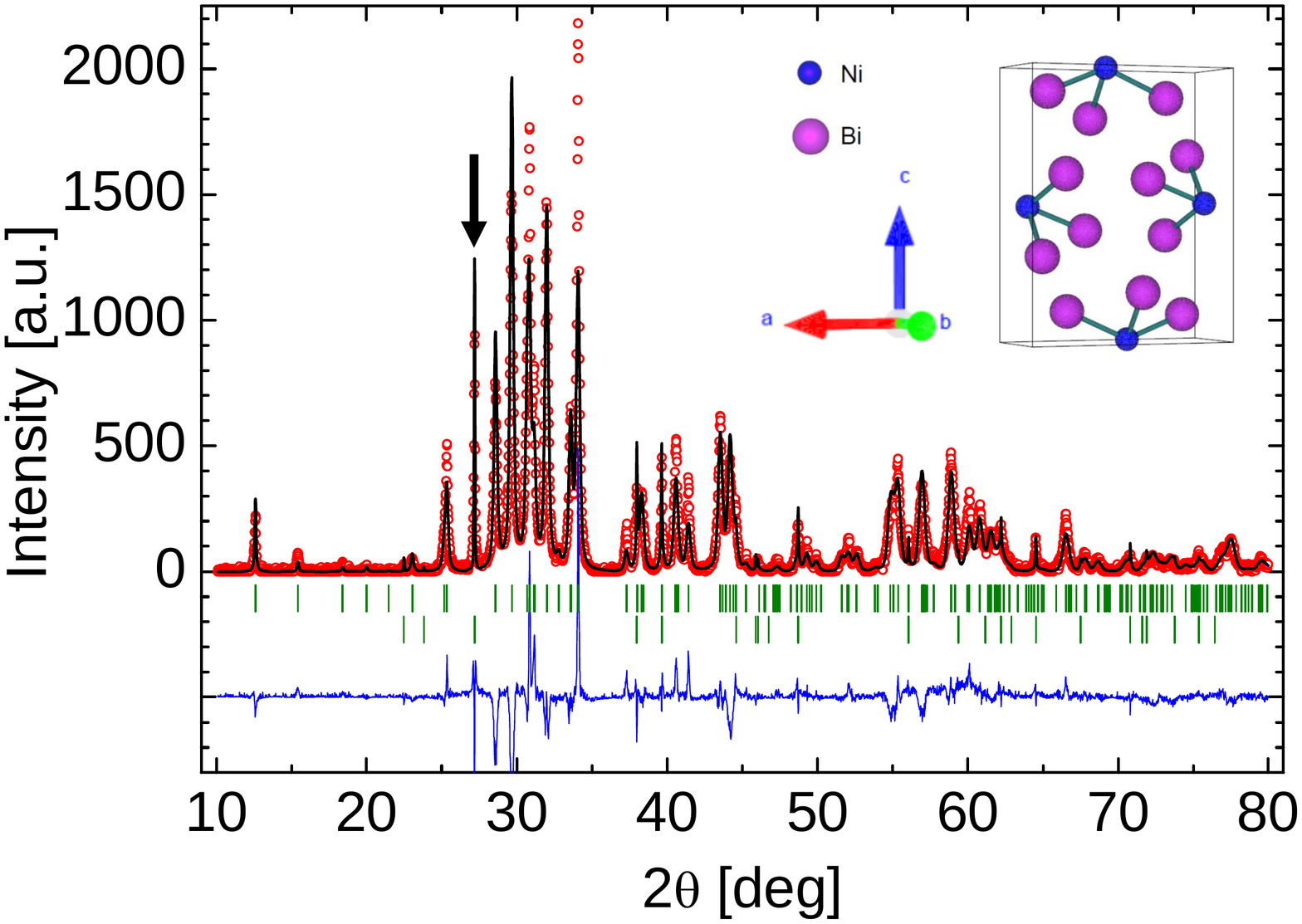}
	\caption{X-ray powder diffraction of NiBi$_3$ needles. Arrows give the peaks associated to NiBi$_3$. Bi flux is also identified by an arrow. In the inset we show the crystal structure of NiBi$_3$.}
\end{figure}

\begin{figure}
	\centering
		\includegraphics[width=0.50\textwidth]{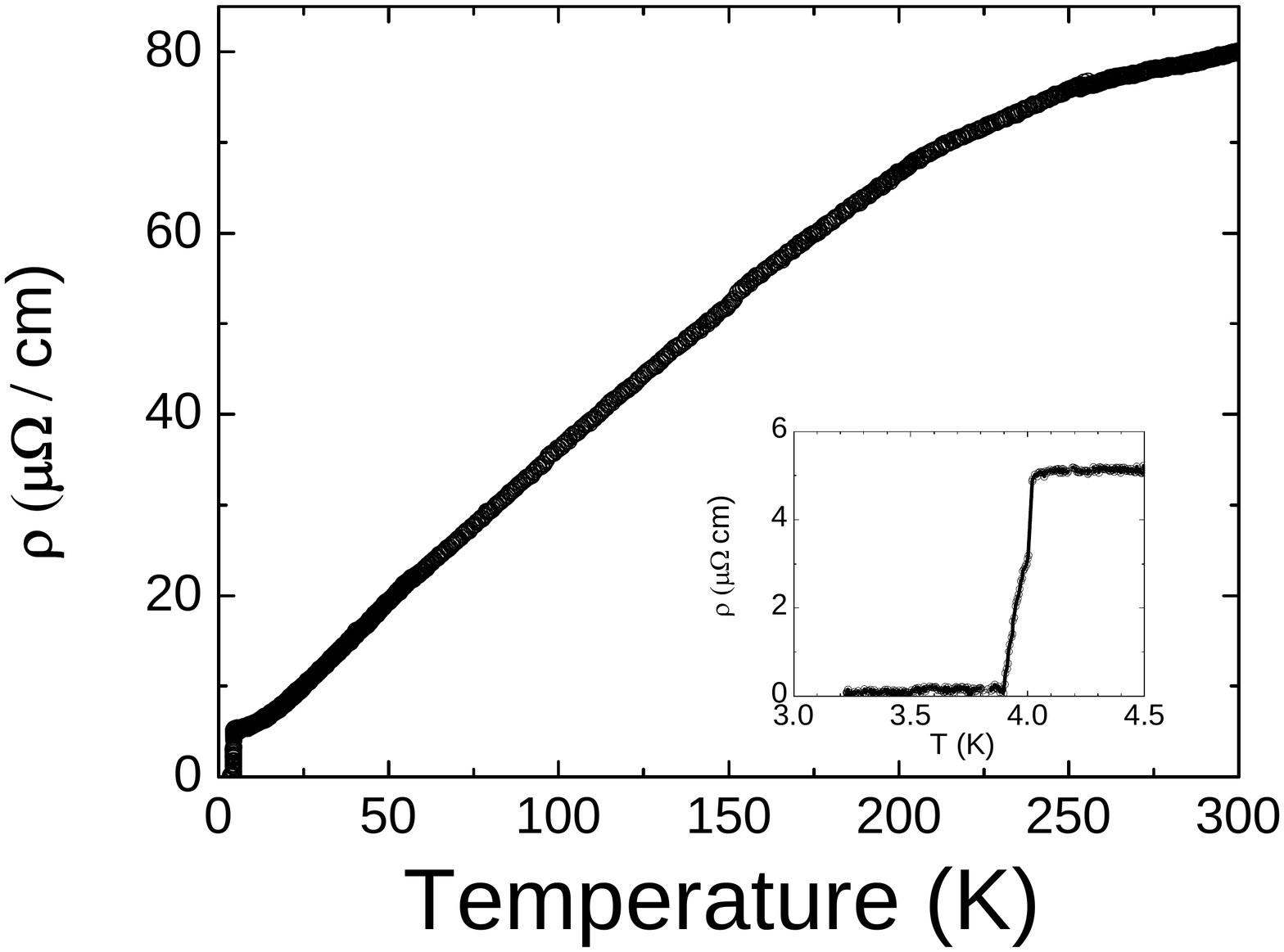}
	\caption{In the main figure we show the temperature dependence of the resistivity between 300 K and 3.5 K. In the lower right inset we zoom into the superconducting transition at the critical temperature.}
\end{figure}

The resistance strongly drops when decreasing temperature, and gives a residual resistance ratio between 4 K and 300 K of 15.3 (Fig.\ 3), indicating good quality single crystal. The superconducting transition starts at 4.0 K with an abrupt drop of the resistivity and ends at 3.9 K, where it becomes zero. Previous resistance and susceptibility measurements give similar residual resistivity values in single crystals\cite{Fujimori2000}. Critical temperature of polycrystals and in nanostructured samples is also similar\cite{Pineiro2011,Zhu2012,Herrmannsdorfer2011,indio}.

\begin{figure}
	\centering
		\includegraphics[width=0.5\textwidth]{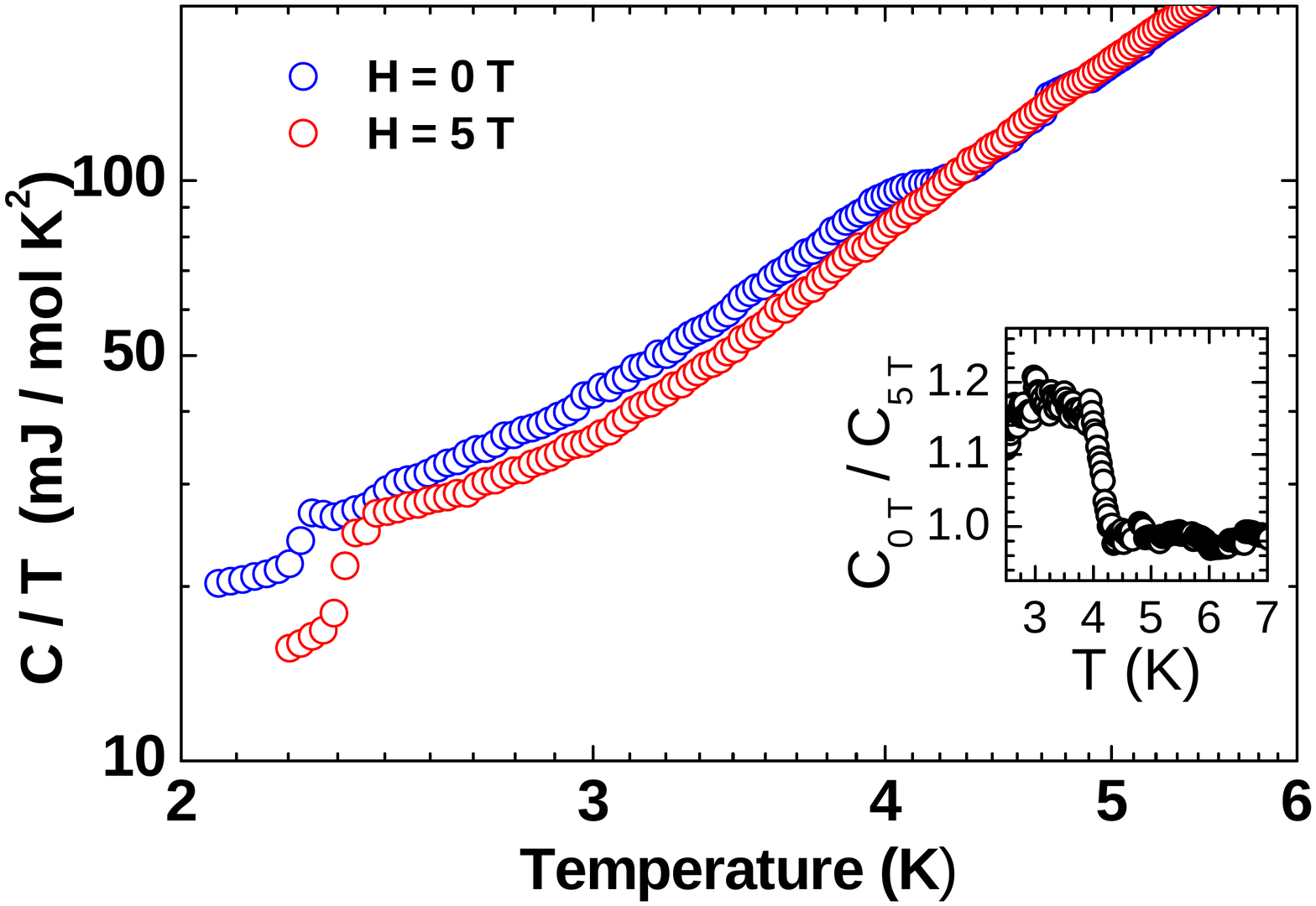}
	\caption{Specific heat divided by temperature $C/T$ as a function of temperature T for NiBi$_3$ at zero magnetic field (red points) and at 5 T (blue points). Inset shows the jump at the superconducting transition as the zero field divided by the high magnetic field specific heat around T$_c$.}
\end{figure}

The specific heat (Fig.\ 4) shows a small peak at the transition, of size expected for a weak coupling BCS superconductor, $\Delta C/T_c=1.43 \gamma$ if we take $\gamma \approx$ 9  mJ/K$^2mol$, which is compatible to the estimated zero temperature extrapolation of $C/T$. The transition is sharp and located at the same temperature as the resistive transition. We find a small anomaly around 2.2 K, which depends of applied magnetic field. The contribution to the specific heat from this anomaly extends to temperatures well above the position of the kink. Actually, in a conventional BCS superconductor, the zero field specific heat should fall below the normal state specific heat approximately around 0.6 $T_c$. Here, the zero field specific heat remains above the 5 T specific heat over the whole temperature range. The observed anomaly at 2.2 K is of magnetic origin and of same order than electronic and phonon contributions. A magnetic transition at this temperature involving the whole sample should give an overwhelming contribution to the specific heat\cite{Canfield95,Durivault03}. Therefore, this anomaly is not due to a full bulk magnetic transition but rather to residual magnetism. No traces of this transition are found in other measurements. The specific heat measurements are the only ones made using pressed powder and not single crystals. Probably, the procedure of crushing the needles into a pellet has brought about a magnetic transition in a small part of the crushed pellet.

The main panel of Fig.\ 5 shows the magnetization of NiBi$_3$ single crystal at 1.8 K. The solid curve corresponds to the first increase of the field after zero field cooling (ZFC) and the dashed line to a decreasing field. From the minimum in the transition between Meissner and Shubnikov states we find $\mu _0$H$_{c1}=$ 12 mT. The normal state is reached at $\mu _0$H$_{c2}=$ 0.35 T indicated by a change in the slope of $M(H)$. The magnetization curves within the superconducting state are rather closed, and we find a reversible behavior over a significant range of magnetic fields. This is expected in a high purity single crystalline sample with low pinning. Single crystalline samples grown by the solid state method give small hysteresis loops \cite{Pineiro2011}, similar to ours. Other samples grown by encapsulation of stoichiometric powder, which show grains and some inhomogeneities, also present larger hysteresis loops\cite{indio}. The magnetization loops with highest hysteresis were for samples grown by a reductive etching of Bi$_{12}$Ni$_4$I$_3$ \cite{Herrmannsdorfer2011}.

\begin{figure}
	\centering
		\includegraphics[width=0.5\textwidth]{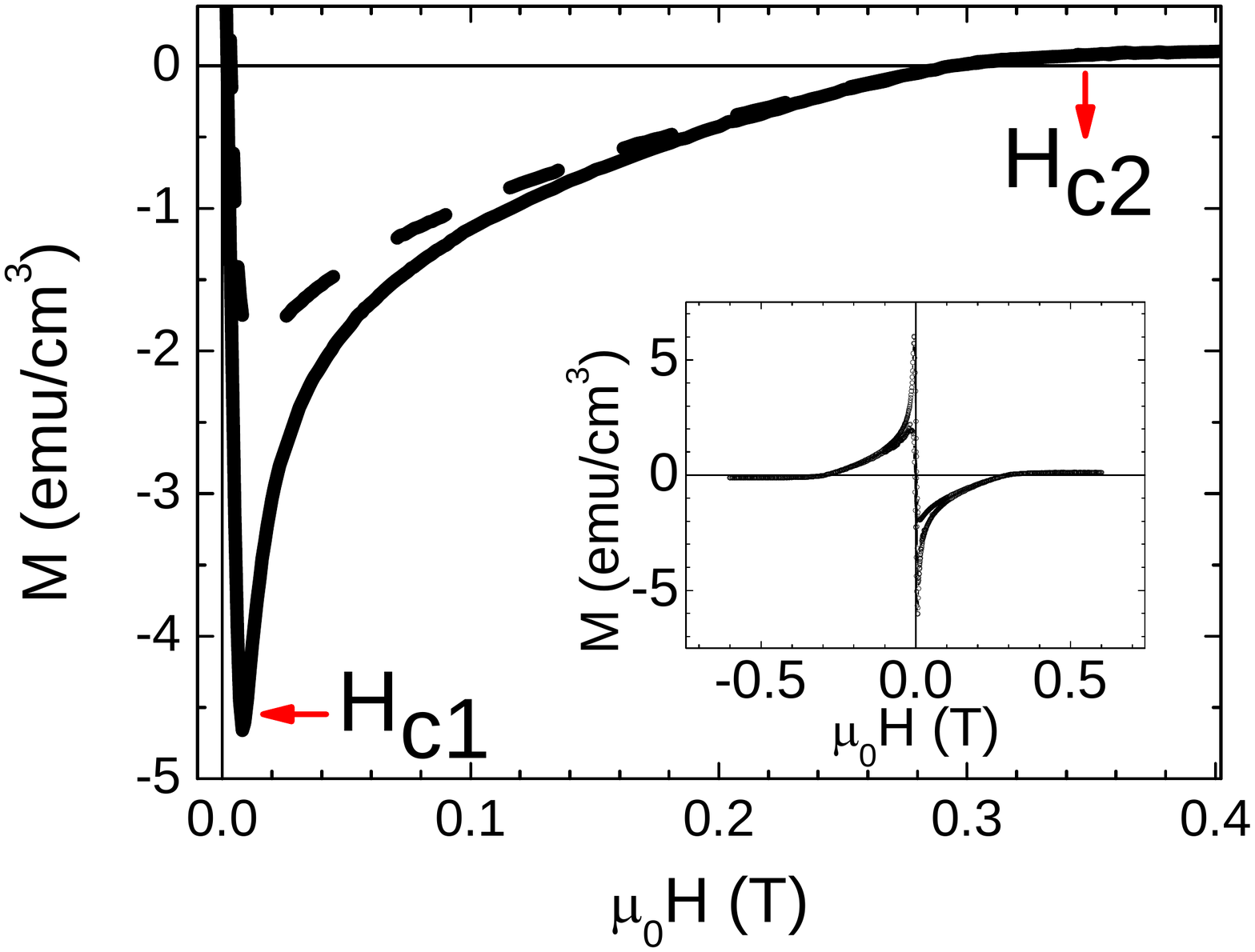}
	\caption{Magnetization as a function of magnetic field at 1.8 K. The continuous line shows increasing field after ZFC. The dashed line shows decreasing field from 6 T. Arrows indicate the superconducting critical fields. Inset shows the full magnetization loop.}
\end{figure}

From $\frac{H_{c2}(0)}{H_{c1}(0)}=\frac{2\kappa^2}{ln(\kappa)},$ (where $H_{c1,2}(0)$ are zero temperature extrapolations of first and second critical fields), $H_{c2}(0)=\frac{\phi _0}{2\pi \mu _0 \xi ^2}$ and $\kappa = \frac{\lambda}{\xi}$, we find $\kappa=$ 5.29 $\pm$ 0.01, $\xi=$ 300 $\pm$ 10 {\AA} and $\lambda=$ 1600 $\pm$ 200 \AA. 
The thermodynamic critical magnetic field is $H_c(0)=$ 50 $\pm$ 3 mT (from $H_{c1}(0)=\frac{H_c(0)}{\sqrt{2}\kappa}ln(\kappa)$).

At magnetic fields above H$_{c2}$, the magnetization shows a peculiar behavior, characterized by a finite magnetization, typical of a ferromagnet. 
 
When we increase the temperature above $T_c$, we can remove the superconducting signal and find the ferromagnetic behavior. It gives hysteresis loops as shown in Fig.\ 6. The saturation field is small, but well defined (M$_S=$ 0.013 emu/g). The loops are rather closed, with a coercitive field of 2.0 mT, and remanence of M$_r=$ 0.0008 emu/g (inset of Fig.\ 6). The saturation magnetization corresponds to $\approx$ 1.74$\cdot$10$^{-3}$ $\mu_{B}$/Ni as compared to the M$_S\approx$ 0.616 $\mu _B$/Ni of elemental ferromagnetic Ni. Thus, magnetism is residual. These values do not change when entering the superconducting phase.

Curves in Fig.\ 5 have been taken after room temperature demagnetazing the ferromagnetic signal. However, if we do not demagnetize, we observe flux trapping in the superconductor by the remanent field. The magnetization curves remain similar as in Fig.\ 5, without additional irreversibility. Thus, the ferromagnetic signal does not produce a strong pinning effect. This is compatible with the soft magnetic features and reminds the situation found in permalloy-superconductor nanostructures\cite{Yang04}.

\begin{figure}
	\centering
		\includegraphics[width=0.5\textwidth]{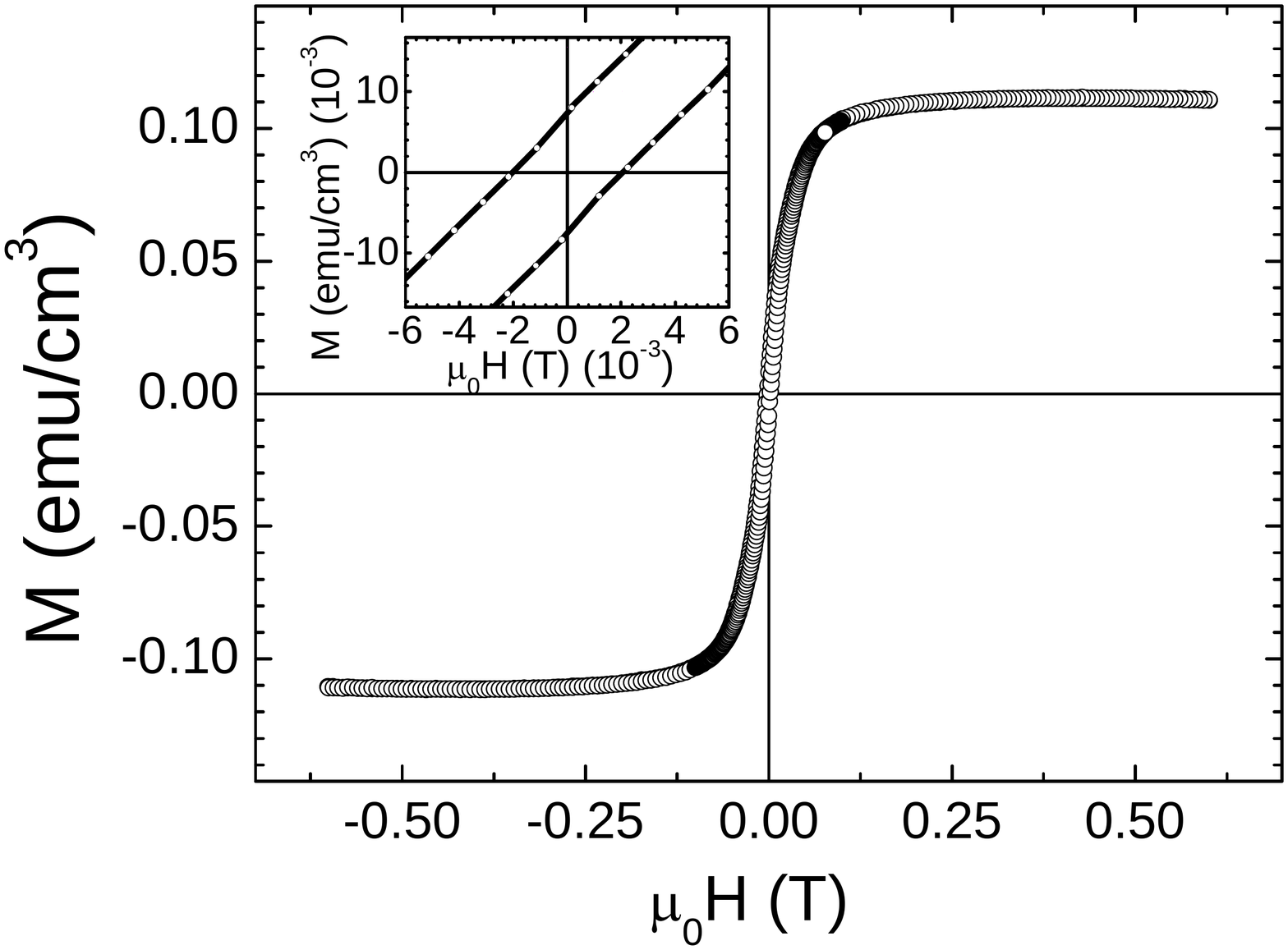}
	\caption{Magnetization hysteresis loop at 10 K. The inset gives a zoom to highlight the behavior close to zero field. Note that the hysteresis is small, indicating soft ferromagnetism.}
\end{figure}

We measured the magnetization of the NiBi$_3$ crystals up to 700 K to search for eventual high-temperature disapperance of the ferromagnetic signal (Fig.\ 7). We find indeed a jump and change of slope at T = 525 K with no further transition until 700 K. The Curie temperature of crystalline Ni is at T$_C$ = 631 K. Thus, the magnetic component is not due to crystalline Ni inclusions. The lack of peaks in X-ray scattering from crystalline Ni implies that there are no crystallized Ni impurities larger that the X-ray coherence lenght of a few 100 \AA. This, of course, does not exclude amorphous Ni. Amorphous Ni is a ferromagnet with a Curie temperature T$_C$ = 530 K. This value is very close to the observed change of slope of Fig.\ 7 \cite{Tamura1969}. We thus conclude that amorphous Ni inclusions produce the observed coexistence of superconductivity and magnetism in this system.

\begin{figure}
	\centering
		\includegraphics[width=0.50\textwidth]{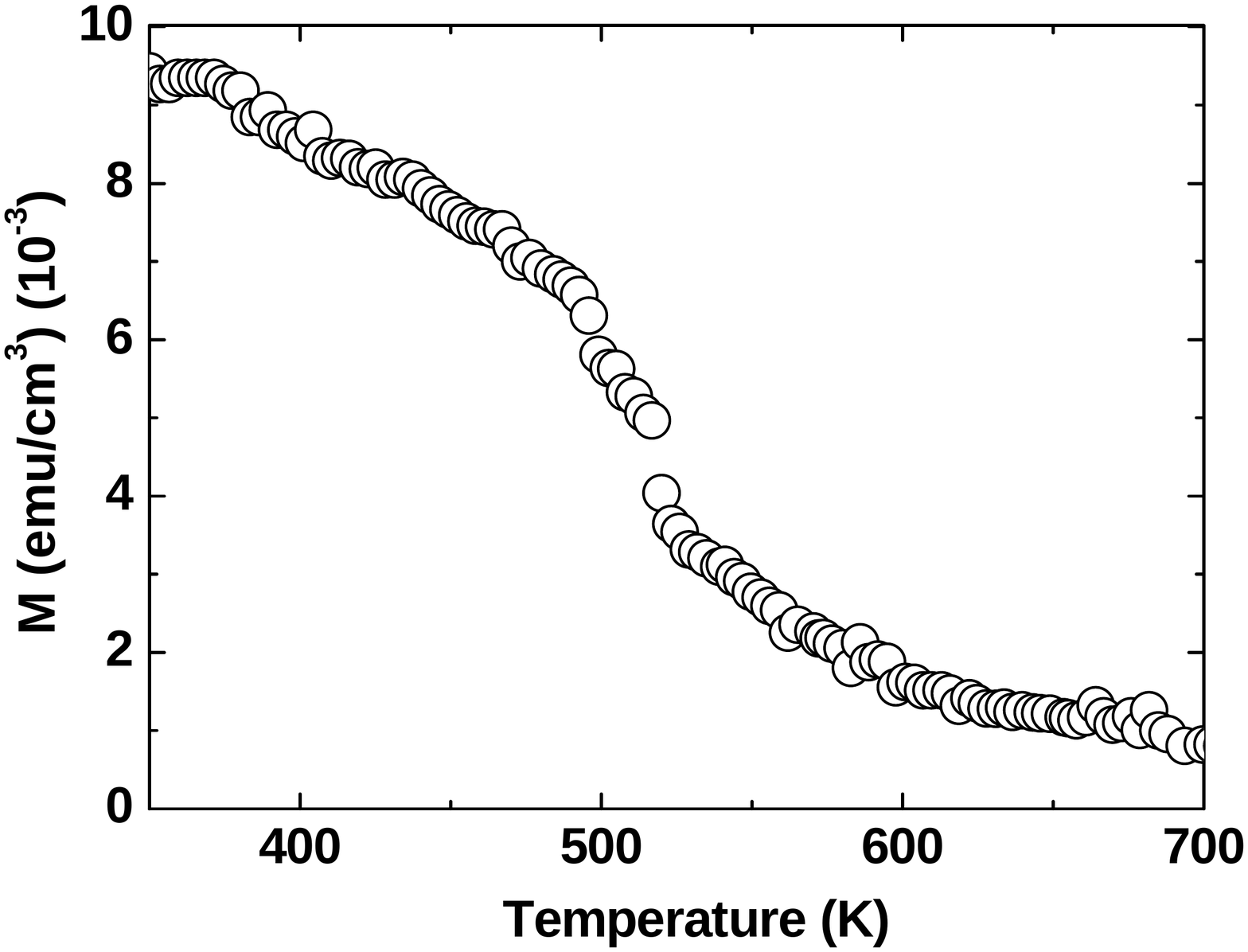}
	\caption{Magnetization of NiBi$_3$\ at high temperatures, in an applied magnetic field of 1 mT. Note the transition at 525 K, close to the Curie temperature of amorphous Ni.}
\end{figure}

\section{Summary and conclusions}

We have prepared NiBi$_3$ single crystal with the flux-growth method, which shows good type II superconducting behavior along with ferromagnetism. We have measured resistivity, specific heat and magnetization. Resistivity shows that we have prepared good quality single crystalline samples, and from specific heat, we obtain a full superconducting transition. In the magnetization curves we find, at the same time, superconducting and ferromagnetic signals. We have shown that the ferromagnetic signal seen in the magnetization experiments stems from amorphous Ni inclusions in the superconducting NiBi$_3$ matrix. Thus, the coexistence of ferromagnetism and superconductivity in this system is extrinsic.

This intermetallic system is prone to give, at the local scale, interesting situations where magnetism and superconductivity show some interplay. This may lead to anomalous proximity effects or vortex pinning features\cite{Buzdin05,Blatter94}. Nanofabrication in form of wires or other structures \cite{Herrmannsdorfer2011} could make use of the likely formation of magnetic inclusions by controling their position and size.

\section{Acknowledgements}

This work was supported by the Spanish MINECO (FIS2011-23488, ACI-2009-0905 and MAT2011-27470-C02-02), by the Comunidad de Madrid through program Nanobiomagnet. We acknowledge P.C. Canfield for teaching us the flux growth method. We also acknowledge Banco Santander for support in setting up the flux growth bench and COST 1201 action.

$^*$ Corresponding author: hermann.suderow@uam.es.


\begin{thebibliography}{41}
\expandafter\ifx\csname natexlab\endcsname\relax\def\natexlab#1{#1}\fi
\expandafter\ifx\csname bibnamefont\endcsname\relax
  \def\bibnamefont#1{#1}\fi
\expandafter\ifx\csname bibfnamefont\endcsname\relax
  \def\bibfnamefont#1{#1}\fi
\expandafter\ifx\csname citenamefont\endcsname\relax
  \def\citenamefont#1{#1}\fi
\expandafter\ifx\csname url\endcsname\relax
  \def\url#1{\texttt{#1}}\fi
\expandafter\ifx\csname urlprefix\endcsname\relax\def\urlprefix{URL }\fi
\providecommand{\bibinfo}[2]{#2}
\providecommand{\eprint}[2][]{\url{#2}}

\bibitem[{\citenamefont{Canfield et~al.}(1998)\citenamefont{Canfield, Gammel,
  and Bishop}}]{Canfield1998}
\bibinfo{author}{\bibfnamefont{P.~C.} \bibnamefont{Canfield}},
  \bibinfo{author}{\bibfnamefont{P.~L.} \bibnamefont{Gammel}},
  \bibnamefont{and} \bibinfo{author}{\bibfnamefont{D.~J.}
  \bibnamefont{Bishop}}, \bibinfo{journal}{Physics Today}
  \textbf{\bibinfo{volume}{51}}, \bibinfo{pages}{40} (\bibinfo{year}{1998}).

\bibitem[{\citenamefont{Kivelson et~al.}(2001)\citenamefont{Kivelson, Aeppli,
  and Emery}}]{Kivelson2001}
\bibinfo{author}{\bibfnamefont{S.~A.} \bibnamefont{Kivelson}},
  \bibinfo{author}{\bibfnamefont{G.}~\bibnamefont{Aeppli}}, \bibnamefont{and}
  \bibinfo{author}{\bibfnamefont{V.~J.} \bibnamefont{Emery}},
  \bibinfo{journal}{Proc. Nat. Acad. Sci.} \textbf{\bibinfo{volume}{98}},
  \bibinfo{pages}{11903} (\bibinfo{year}{2001}).

\bibitem[{\citenamefont{Aoki et~al.}(2001)\citenamefont{Aoki, Huxley,
  Ressouche, Braithwaite, Flouquet, Brison, Lhotel, and Paulsen}}]{Aoki2001}
\bibinfo{author}{\bibfnamefont{D.}~\bibnamefont{Aoki}},
  \bibinfo{author}{\bibfnamefont{A.}~\bibnamefont{Huxley}},
  \bibinfo{author}{\bibfnamefont{E.}~\bibnamefont{Ressouche}},
  \bibinfo{author}{\bibfnamefont{D.}~\bibnamefont{Braithwaite}},
  \bibinfo{author}{\bibfnamefont{J.}~\bibnamefont{Flouquet}},
  \bibinfo{author}{\bibfnamefont{J.-P.} \bibnamefont{Brison}},
  \bibinfo{author}{\bibfnamefont{E.}~\bibnamefont{Lhotel}}, \bibnamefont{and}
  \bibinfo{author}{\bibfnamefont{C.}~\bibnamefont{Paulsen}},
  \bibinfo{journal}{Nature} \textbf{\bibinfo{volume}{413}},
  \bibinfo{pages}{613} (\bibinfo{year}{2001}).

\bibitem[{\citenamefont{Coronado et~al.}(2010)\citenamefont{Coronado,
  Marti-Gastaldo, Navarro-Moratalla, Ribera, Blundell, and Baker}}]{Coronado10}
\bibinfo{author}{\bibfnamefont{E.}~\bibnamefont{Coronado}},
  \bibinfo{author}{\bibfnamefont{C.}~\bibnamefont{Marti-Gastaldo}},
  \bibinfo{author}{\bibfnamefont{E.}~\bibnamefont{Navarro-Moratalla}},
  \bibinfo{author}{\bibfnamefont{A.}~\bibnamefont{Ribera}},
  \bibinfo{author}{\bibfnamefont{S.}~\bibnamefont{Blundell}}, \bibnamefont{and}
  \bibinfo{author}{\bibfnamefont{P.}~\bibnamefont{Baker}},
  \bibinfo{journal}{Nat. Chem.} \textbf{\bibinfo{volume}{2}},
  \bibinfo{pages}{1031} (\bibinfo{year}{2010}).

\bibitem[{\citenamefont{Matthias}(1953)}]{Matthias1953}
\bibinfo{author}{\bibfnamefont{B.~T.} \bibnamefont{Matthias}},
  \bibinfo{journal}{Phys. Rev.} \textbf{\bibinfo{volume}{92}},
  \bibinfo{pages}{874} (\bibinfo{year}{1953}).

\bibitem[{\citenamefont{Chervenak and Valles}(1995)}]{Chervenak1995}
\bibinfo{author}{\bibfnamefont{J.~A.} \bibnamefont{Chervenak}}
  \bibnamefont{and} \bibinfo{author}{\bibfnamefont{J.~M.}
  \bibnamefont{Valles}}, \bibinfo{journal}{Phys. Rev. B}
  \textbf{\bibinfo{volume}{51}}, \bibinfo{pages}{11977} (\bibinfo{year}{1995}).

\bibitem[{\citenamefont{Garrett et~al.}(1998)\citenamefont{Garrett, Beckmann,
  and Bergmann}}]{Garret1998}
\bibinfo{author}{\bibfnamefont{D.}~\bibnamefont{Garrett}},
  \bibinfo{author}{\bibfnamefont{H.}~\bibnamefont{Beckmann}}, \bibnamefont{and}
  \bibinfo{author}{\bibfnamefont{G.}~\bibnamefont{Bergmann}},
  \bibinfo{journal}{Phys. Rev. B} \textbf{\bibinfo{volume}{57}},
  \bibinfo{pages}{2732} (\bibinfo{year}{1998}).

\bibitem[{\citenamefont{Smith and Ambegaokar}(2000)}]{Smith2000}
\bibinfo{author}{\bibfnamefont{R.~A.} \bibnamefont{Smith}} \bibnamefont{and}
  \bibinfo{author}{\bibfnamefont{V.}~\bibnamefont{Ambegaokar}},
  \bibinfo{journal}{Phys. Rev. B} \textbf{\bibinfo{volume}{62}},
  \bibinfo{pages}{5913} (\bibinfo{year}{2000}).

\bibitem[{\citenamefont{Flouquet and Buzdin}(2002)}]{flouquet2002ferromagnetic}
\bibinfo{author}{\bibfnamefont{J.}~\bibnamefont{Flouquet}} \bibnamefont{and}
  \bibinfo{author}{\bibfnamefont{A.}~\bibnamefont{Buzdin}},
  \bibinfo{journal}{Physics World} \textbf{\bibinfo{volume}{15}},
  \bibinfo{pages}{41} (\bibinfo{year}{2002}).

\bibitem[{\citenamefont{Bulaevskii et~al.}(1985)\citenamefont{Bulaevskii,
  Buzdin, Kuli\'c, and Panjukov}}]{Bulaevski85}
\bibinfo{author}{\bibfnamefont{L.~N.} \bibnamefont{Bulaevskii}},
  \bibinfo{author}{\bibfnamefont{A.~I.} \bibnamefont{Buzdin}},
  \bibinfo{author}{\bibfnamefont{M.~L.} \bibnamefont{Kuli\'c}},
  \bibnamefont{and} \bibinfo{author}{\bibfnamefont{S.~V.}
  \bibnamefont{Panjukov}}, \bibinfo{journal}{Advances in Physics}
  \textbf{\bibinfo{volume}{34}}, \bibinfo{pages}{175 } (\bibinfo{year}{1985}).

\bibitem[{\citenamefont{Fertig et~al.}(1977)\citenamefont{Fertig, Johnston,
  DeLong, McCallum, Maple, and Matthias}}]{Fertig1977}
\bibinfo{author}{\bibfnamefont{W.~A.} \bibnamefont{Fertig}},
  \bibinfo{author}{\bibfnamefont{D.~C.} \bibnamefont{Johnston}},
  \bibinfo{author}{\bibfnamefont{L.~E.} \bibnamefont{DeLong}},
  \bibinfo{author}{\bibfnamefont{R.~W.} \bibnamefont{McCallum}},
  \bibinfo{author}{\bibfnamefont{M.~B.} \bibnamefont{Maple}}, \bibnamefont{and}
  \bibinfo{author}{\bibfnamefont{B.~T.} \bibnamefont{Matthias}},
  \bibinfo{journal}{Phys. Rev. Lett.} \textbf{\bibinfo{volume}{38}},
  \bibinfo{pages}{987} (\bibinfo{year}{1977}).

\bibitem[{\citenamefont{Ott et~al.}(1978)\citenamefont{Ott, Fertig, Johnston,
  Maple, and Matthias}}]{Ott1978}
\bibinfo{author}{\bibfnamefont{H.~R.} \bibnamefont{Ott}},
  \bibinfo{author}{\bibfnamefont{W.~A.} \bibnamefont{Fertig}},
  \bibinfo{author}{\bibfnamefont{D.~C.} \bibnamefont{Johnston}},
  \bibinfo{author}{\bibfnamefont{M.~B.} \bibnamefont{Maple}}, \bibnamefont{and}
  \bibinfo{author}{\bibfnamefont{B.~T.} \bibnamefont{Matthias}},
  \bibinfo{journal}{J. Low Temp. Phys.} \textbf{\bibinfo{volume}{33}},
  \bibinfo{pages}{159} (\bibinfo{year}{1978}).

\bibitem[{\citenamefont{Lynn et~al.}(1985)\citenamefont{Lynn, Gotaas, Shelton,
  Horng, and Glinka}}]{Lynn1985}
\bibinfo{author}{\bibfnamefont{J.~W.} \bibnamefont{Lynn}},
  \bibinfo{author}{\bibfnamefont{J.~A.} \bibnamefont{Gotaas}},
  \bibinfo{author}{\bibfnamefont{R.~N.} \bibnamefont{Shelton}},
  \bibinfo{author}{\bibfnamefont{H.~E.} \bibnamefont{Horng}}, \bibnamefont{and}
  \bibinfo{author}{\bibfnamefont{C.~J.} \bibnamefont{Glinka}},
  \bibinfo{journal}{Phys. Rev. B} \textbf{\bibinfo{volume}{31}},
  \bibinfo{pages}{5756} (\bibinfo{year}{1985}).

\bibitem[{\citenamefont{Prozorov et~al.}(2008)\citenamefont{Prozorov, Vannette,
  Law, Bud'ko, and Canfield}}]{Prozorov2008}
\bibinfo{author}{\bibfnamefont{R.}~\bibnamefont{Prozorov}},
  \bibinfo{author}{\bibfnamefont{M.~D.} \bibnamefont{Vannette}},
  \bibinfo{author}{\bibfnamefont{S.~A.} \bibnamefont{Law}},
  \bibinfo{author}{\bibfnamefont{S.~L.} \bibnamefont{Bud'ko}},
  \bibnamefont{and} \bibinfo{author}{\bibfnamefont{P.~C.}
  \bibnamefont{Canfield}}, \bibinfo{journal}{Phys. Rev. B}
  \textbf{\bibinfo{volume}{77}}, \bibinfo{pages}{100503}
  (\bibinfo{year}{2008}).

\bibitem[{\citenamefont{Crespo et~al.}(2009)\citenamefont{Crespo, Rodrigo,
  Suderow, Vieira, Hinks, and Schuller}}]{PhysRevLett.102.237002}
\bibinfo{author}{\bibfnamefont{V.}~\bibnamefont{Crespo}},
  \bibinfo{author}{\bibfnamefont{J.~G.} \bibnamefont{Rodrigo}},
  \bibinfo{author}{\bibfnamefont{H.}~\bibnamefont{Suderow}},
  \bibinfo{author}{\bibfnamefont{S.}~\bibnamefont{Vieira}},
  \bibinfo{author}{\bibfnamefont{D.~G.} \bibnamefont{Hinks}}, \bibnamefont{and}
  \bibinfo{author}{\bibfnamefont{I.~K.} \bibnamefont{Schuller}},
  \bibinfo{journal}{Phys. Rev. Lett.} \textbf{\bibinfo{volume}{102}},
  \bibinfo{pages}{237002} (\bibinfo{year}{2009}).

\bibitem[{\citenamefont{Crespo et~al.}(2006{\natexlab{a}})\citenamefont{Crespo,
  Suderow, Vieira, Bud'ko, and Canfield}}]{PhysRevLett.96.027003}
\bibinfo{author}{\bibfnamefont{M.}~\bibnamefont{Crespo}},
  \bibinfo{author}{\bibfnamefont{H.}~\bibnamefont{Suderow}},
  \bibinfo{author}{\bibfnamefont{S.}~\bibnamefont{Vieira}},
  \bibinfo{author}{\bibfnamefont{S.}~\bibnamefont{Bud'ko}}, \bibnamefont{and}
  \bibinfo{author}{\bibfnamefont{P.~C.} \bibnamefont{Canfield}},
  \bibinfo{journal}{Phys. Rev. Lett.} \textbf{\bibinfo{volume}{96}},
  \bibinfo{pages}{027003} (\bibinfo{year}{2006}{\natexlab{a}}).

\bibitem[{\citenamefont{Galvis et~al.}(2012)\citenamefont{Galvis, Crespo,
  Guillam\'on, Suderow, Vieira, Hern\'andez, Bud'ko, and
  Canfield}}]{Galvis20121076}
\bibinfo{author}{\bibfnamefont{J.}~\bibnamefont{Galvis}},
  \bibinfo{author}{\bibfnamefont{M.}~\bibnamefont{Crespo}},
  \bibinfo{author}{\bibfnamefont{I.}~\bibnamefont{Guillam\'on}},
  \bibinfo{author}{\bibfnamefont{H.}~\bibnamefont{Suderow}},
  \bibinfo{author}{\bibfnamefont{S.}~\bibnamefont{Vieira}},
  \bibinfo{author}{\bibfnamefont{M.~G.} \bibnamefont{Hern\'andez}},
  \bibinfo{author}{\bibfnamefont{S.}~\bibnamefont{Bud'ko}}, \bibnamefont{and}
  \bibinfo{author}{\bibfnamefont{P.}~\bibnamefont{Canfield}},
  \bibinfo{journal}{Solid State Communications} \textbf{\bibinfo{volume}{152}},
  \bibinfo{pages}{1076 } (\bibinfo{year}{2012}).

\bibitem[{\citenamefont{Crespo et~al.}(2006{\natexlab{b}})\citenamefont{Crespo,
  Suderow, Vieira, Bud'ko, and Canfield}}]{Crespo2006471}
\bibinfo{author}{\bibfnamefont{M.}~\bibnamefont{Crespo}},
  \bibinfo{author}{\bibfnamefont{H.}~\bibnamefont{Suderow}},
  \bibinfo{author}{\bibfnamefont{S.}~\bibnamefont{Vieira}},
  \bibinfo{author}{\bibfnamefont{S.}~\bibnamefont{Bud'ko}}, \bibnamefont{and}
  \bibinfo{author}{\bibfnamefont{P.~C.} \bibnamefont{Canfield}},
  \bibinfo{journal}{Physica B: Condensed Matter}
  \textbf{\bibinfo{volume}{378-80}}, \bibinfo{pages}{471 }
  (\bibinfo{year}{2006}{\natexlab{b}}).

\bibitem[{\citenamefont{Maple}(1986)}]{Maple1986}
\bibinfo{author}{\bibfnamefont{M.~B.} \bibnamefont{Maple}},
  \bibinfo{journal}{Physics Today} \textbf{\bibinfo{volume}{39}},
  \bibinfo{pages}{72} (\bibinfo{year}{1986}).

\bibitem[{\citenamefont{Sangiao et~al.}(2011)\citenamefont{Sangiao, De~Teresa,
  Ibarra, Guillam\'on, Suderow, Vieira, and Morell\'on}}]{Sangiao11}
\bibinfo{author}{\bibfnamefont{S.}~\bibnamefont{Sangiao}},
  \bibinfo{author}{\bibfnamefont{J.~M.} \bibnamefont{De~Teresa}},
  \bibinfo{author}{\bibfnamefont{M.~R.} \bibnamefont{Ibarra}},
  \bibinfo{author}{\bibfnamefont{I.}~\bibnamefont{Guillam\'on}},
  \bibinfo{author}{\bibfnamefont{H.}~\bibnamefont{Suderow}},
  \bibinfo{author}{\bibfnamefont{S.}~\bibnamefont{Vieira}}, \bibnamefont{and}
  \bibinfo{author}{\bibfnamefont{L.}~\bibnamefont{Morell\'on}},
  \bibinfo{journal}{Phys. Rev. B} \textbf{\bibinfo{volume}{84}},
  \bibinfo{pages}{233402} (\bibinfo{year}{2011}).

\bibitem[{\citenamefont{Guillamon et~al.}(2010)\citenamefont{Guillamon, Crespo,
  Suderow, Vieira, Brison, Bud'ko, and Canfield}}]{Guillamon2010771}
\bibinfo{author}{\bibfnamefont{I.}~\bibnamefont{Guillamon}},
  \bibinfo{author}{\bibfnamefont{M.}~\bibnamefont{Crespo}},
  \bibinfo{author}{\bibfnamefont{H.}~\bibnamefont{Suderow}},
  \bibinfo{author}{\bibfnamefont{S.}~\bibnamefont{Vieira}},
  \bibinfo{author}{\bibfnamefont{J.~P.} \bibnamefont{Brison}},
  \bibinfo{author}{\bibfnamefont{S.~L.} \bibnamefont{Bud'ko}},
  \bibnamefont{and} \bibinfo{author}{\bibfnamefont{P.~C.}
  \bibnamefont{Canfield}}, \bibinfo{journal}{Physica C}
  \textbf{\bibinfo{volume}{470}}, \bibinfo{pages}{771 } (\bibinfo{year}{2010}).

\bibitem[{\citenamefont{Suderow et~al.}(2001)\citenamefont{Suderow,
  Martinez-Samper, Luchier, Brison, Vieira, and Canfield}}]{Suderow01}
\bibinfo{author}{\bibfnamefont{H.}~\bibnamefont{Suderow}},
  \bibinfo{author}{\bibfnamefont{P.}~\bibnamefont{Martinez-Samper}},
  \bibinfo{author}{\bibfnamefont{N.}~\bibnamefont{Luchier}},
  \bibinfo{author}{\bibfnamefont{J.~P.} \bibnamefont{Brison}},
  \bibinfo{author}{\bibfnamefont{S.}~\bibnamefont{Vieira}}, \bibnamefont{and}
  \bibinfo{author}{\bibfnamefont{P.~C.} \bibnamefont{Canfield}},
  \bibinfo{journal}{Phys. Rev. B} \textbf{\bibinfo{volume}{64}},
  \bibinfo{pages}{020503} (\bibinfo{year}{2001}).

\bibitem[{\citenamefont{Cordoba and et~al.}(2013)}]{Cordoba13}
\bibinfo{author}{\bibfnamefont{R.}~\bibnamefont{Cordoba}} \bibnamefont{and}
  \bibinfo{author}{\bibnamefont{et~al.}}, \bibinfo{journal}{Nat. Comm.}
  \textbf{\bibinfo{volume}{4}}, \bibinfo{pages}{1437} (\bibinfo{year}{2013}).

\bibitem[{\citenamefont{Buzdin}(2005)}]{Buzdin05}
\bibinfo{author}{\bibfnamefont{A.~I.} \bibnamefont{Buzdin}},
  \bibinfo{journal}{Rev. Mod. Phys.} \textbf{\bibinfo{volume}{77}},
  \bibinfo{pages}{935} (\bibinfo{year}{2005}).

\bibitem[{\citenamefont{Steglich et~al.}(1979)\citenamefont{Steglich, Aarts,
  Bredl, Lieke, Meschede, Franz, and Sch{\"a}fer}}]{Steglich1979}
\bibinfo{author}{\bibfnamefont{F.}~\bibnamefont{Steglich}},
  \bibinfo{author}{\bibfnamefont{J.}~\bibnamefont{Aarts}},
  \bibinfo{author}{\bibfnamefont{C.~D.} \bibnamefont{Bredl}},
  \bibinfo{author}{\bibfnamefont{W.}~\bibnamefont{Lieke}},
  \bibinfo{author}{\bibfnamefont{D.}~\bibnamefont{Meschede}},
  \bibinfo{author}{\bibfnamefont{W.}~\bibnamefont{Franz}}, \bibnamefont{and}
  \bibinfo{author}{\bibfnamefont{H.}~\bibnamefont{Sch{\"a}fer}},
  \bibinfo{journal}{Phys. Rev. Lett.} \textbf{\bibinfo{volume}{43}},
  \bibinfo{pages}{1892} (\bibinfo{year}{1979}).

\bibitem[{\citenamefont{Brison et~al.}(2000)\citenamefont{Brison, Gl{\'e}mot,
  Suderow, Huxley, Kambe, and Flouquet}}]{Brison2000}
\bibinfo{author}{\bibfnamefont{J.-P.} \bibnamefont{Brison}},
  \bibinfo{author}{\bibfnamefont{L.}~\bibnamefont{Gl{\'e}mot}},
  \bibinfo{author}{\bibfnamefont{H.}~\bibnamefont{Suderow}},
  \bibinfo{author}{\bibfnamefont{A.}~\bibnamefont{Huxley}},
  \bibinfo{author}{\bibfnamefont{S.}~\bibnamefont{Kambe}}, \bibnamefont{and}
  \bibinfo{author}{\bibfnamefont{J.}~\bibnamefont{Flouquet}},
  \bibinfo{journal}{Phys. B} \textbf{\bibinfo{volume}{280}},
  \bibinfo{pages}{165} (\bibinfo{year}{2000}).

\bibitem[{\citenamefont{Steglich}(2005)}]{Steglich2005}
\bibinfo{author}{\bibfnamefont{F.}~\bibnamefont{Steglich}},
  \bibinfo{journal}{Phys. B} \textbf{\bibinfo{volume}{359-61}},
  \bibinfo{pages}{326} (\bibinfo{year}{2005}).

\bibitem[{\citenamefont{Steglich et~al.}(2012)\citenamefont{Steglich, Stockert,
  Wirth, Geibel, Yuan, Kirchner, and Si}}]{Steglich2012}
\bibinfo{author}{\bibfnamefont{F.}~\bibnamefont{Steglich}},
  \bibinfo{author}{\bibfnamefont{O.}~\bibnamefont{Stockert}},
  \bibinfo{author}{\bibfnamefont{S.}~\bibnamefont{Wirth}},
  \bibinfo{author}{\bibfnamefont{C.}~\bibnamefont{Geibel}},
  \bibinfo{author}{\bibfnamefont{H.~Q.} \bibnamefont{Yuan}},
  \bibinfo{author}{\bibfnamefont{S.}~\bibnamefont{Kirchner}}, \bibnamefont{and}
  \bibinfo{author}{\bibfnamefont{Q.}~\bibnamefont{Si}},
  \bibinfo{journal}{arXiv:1208.3684}  (\bibinfo{year}{2012}).

\bibitem[{\citenamefont{Fujimori et~al.}(2000)\citenamefont{Fujimori, Kan,
  Shinozaki, and Kawaguti}}]{Fujimori2000}
\bibinfo{author}{\bibfnamefont{Y.}~\bibnamefont{Fujimori}},
  \bibinfo{author}{\bibfnamefont{S.}~\bibnamefont{Kan}},
  \bibinfo{author}{\bibfnamefont{B.}~\bibnamefont{Shinozaki}},
  \bibnamefont{and} \bibinfo{author}{\bibfnamefont{T.}~\bibnamefont{Kawaguti}},
  \bibinfo{journal}{J. Phys. Soc. Jpn.} \textbf{\bibinfo{volume}{69}},
  \bibinfo{pages}{3017} (\bibinfo{year}{2000}).

\bibitem[{\citenamefont{Martinez~Pi{\~n}eiro
  et~al.}(2011)\citenamefont{Martinez~Pi{\~n}eiro, Ruiz~Herrera, Escudero, and
  Bucio}}]{Pineiro2011}
\bibinfo{author}{\bibfnamefont{E.~L.} \bibnamefont{Martinez~Pi{\~n}eiro}},
  \bibinfo{author}{\bibfnamefont{B.~L.} \bibnamefont{Ruiz~Herrera}},
  \bibinfo{author}{\bibfnamefont{R.}~\bibnamefont{Escudero}}, \bibnamefont{and}
  \bibinfo{author}{\bibfnamefont{L.}~\bibnamefont{Bucio}},
  \bibinfo{journal}{Solid State Commun.} \textbf{\bibinfo{volume}{151}},
  \bibinfo{pages}{425} (\bibinfo{year}{2011}).

\bibitem[{\citenamefont{Zhu et~al.}(2012)\citenamefont{Zhu, Lei, Petrovic, and
  Zhang}}]{Zhu2012}
\bibinfo{author}{\bibfnamefont{X.}~\bibnamefont{Zhu}},
  \bibinfo{author}{\bibfnamefont{H.}~\bibnamefont{Lei}},
  \bibinfo{author}{\bibfnamefont{C.}~\bibnamefont{Petrovic}}, \bibnamefont{and}
  \bibinfo{author}{\bibfnamefont{Y.}~\bibnamefont{Zhang}},
  \bibinfo{journal}{Phys. Rev. B} \textbf{\bibinfo{volume}{86}},
  \bibinfo{pages}{024527} (\bibinfo{year}{2012}).

\bibitem[{\citenamefont{Herrmannsd{\"o}rfer
  et~al.}(2011)\citenamefont{Herrmannsd{\"o}rfer, Skrotzki, Wosnitza,
  K{\"o}hler, Boldt, and Ruck}}]{Herrmannsdorfer2011}
\bibinfo{author}{\bibfnamefont{T.}~\bibnamefont{Herrmannsd{\"o}rfer}},
  \bibinfo{author}{\bibfnamefont{R.}~\bibnamefont{Skrotzki}},
  \bibinfo{author}{\bibfnamefont{J.}~\bibnamefont{Wosnitza}},
  \bibinfo{author}{\bibfnamefont{D.}~\bibnamefont{K{\"o}hler}},
  \bibinfo{author}{\bibfnamefont{R.}~\bibnamefont{Boldt}}, \bibnamefont{and}
  \bibinfo{author}{\bibfnamefont{M.}~\bibnamefont{Ruck}},
  \bibinfo{journal}{Phys. Rev. B} \textbf{\bibinfo{volume}{83}},
  \bibinfo{pages}{140501} (\bibinfo{year}{2011}).

\bibitem[{\citenamefont{Canfield and Fisk}(1992)}]{Canfield1992}
\bibinfo{author}{\bibfnamefont{P.~C.} \bibnamefont{Canfield}} \bibnamefont{and}
  \bibinfo{author}{\bibfnamefont{Z.}~\bibnamefont{Fisk}},
  \bibinfo{journal}{Phil. Mag. B} \textbf{\bibinfo{volume}{65}},
  \bibinfo{pages}{1117} (\bibinfo{year}{1992}).

\bibitem[{\citenamefont{Canfield et~al.}(1995)\citenamefont{Canfield, Cho, and
  Dennis}}]{Canfield95}
\bibinfo{author}{\bibfnamefont{P.~C.} \bibnamefont{Canfield}},
  \bibinfo{author}{\bibfnamefont{B.~K.} \bibnamefont{Cho}}, \bibnamefont{and}
  \bibinfo{author}{\bibfnamefont{K.~W.} \bibnamefont{Dennis}},
  \bibinfo{journal}{Physica B} \textbf{\bibinfo{volume}{215}},
  \bibinfo{pages}{337 } (\bibinfo{year}{1995}).

\bibitem[{\citenamefont{Canfield}(2009)}]{CanfieldBook}
\bibinfo{author}{\bibfnamefont{P.~C.} \bibnamefont{Canfield}},
  \emph{\bibinfo{title}{Solution growth of intermetallic single crystals: a
  beginner's guide.}} (\bibinfo{year}{2009}), chap.~\bibinfo{chapter}{2}, pp.
  \bibinfo{pages}{93--111}.

\bibitem[{\citenamefont{Ruck and S\"ohnel}(2006)}]{Ruck06}
\bibinfo{author}{\bibfnamefont{M.}~\bibnamefont{Ruck}} \bibnamefont{and}
  \bibinfo{author}{\bibfnamefont{T.~Z.} \bibnamefont{S\"ohnel}},
  \bibinfo{journal}{Z. Naturforsch. B: Chem. Sci.}
  \textbf{\bibinfo{volume}{61}}, \bibinfo{pages}{785} (\bibinfo{year}{2006}).

\bibitem[{\citenamefont{Kumar et~al.}(2011)\citenamefont{Kumar, Kumar,
  Vajpayee, Gahtori, Sharma, Ahluwalia, Auluck, and Awana.}}]{indio}
\bibinfo{author}{\bibfnamefont{J.}~\bibnamefont{Kumar}},
  \bibinfo{author}{\bibfnamefont{A.}~\bibnamefont{Kumar}},
  \bibinfo{author}{\bibfnamefont{A.}~\bibnamefont{Vajpayee}},
  \bibinfo{author}{\bibfnamefont{B.}~\bibnamefont{Gahtori}},
  \bibinfo{author}{\bibfnamefont{D.}~\bibnamefont{Sharma}},
  \bibinfo{author}{\bibfnamefont{P.~K.} \bibnamefont{Ahluwalia}},
  \bibinfo{author}{\bibfnamefont{S.}~\bibnamefont{Auluck}}, \bibnamefont{and}
  \bibinfo{author}{\bibfnamefont{V.~P.~S.} \bibnamefont{Awana.}},
  \bibinfo{journal}{Supercond. Sci. Technol} \textbf{\bibinfo{volume}{24}},
  \bibinfo{pages}{085002} (\bibinfo{year}{2011}).

\bibitem[{\citenamefont{Durivault and et~al.}(2003)}]{Durivault03}
\bibinfo{author}{\bibfnamefont{L.}~\bibnamefont{Durivault}} \bibnamefont{and}
  \bibinfo{author}{\bibnamefont{et~al.}}, \bibinfo{journal}{J. Phys. Cond.
  Matt.} \textbf{\bibinfo{volume}{15}}, \bibinfo{pages}{77}
  (\bibinfo{year}{2003}).

\bibitem[{\citenamefont{Yang et~al.}(2004)\citenamefont{Yang, Lange, Volodin,
  Szymczak, and Moshchalkov}}]{Yang04}
\bibinfo{author}{\bibfnamefont{Z.}~\bibnamefont{Yang}},
  \bibinfo{author}{\bibfnamefont{M.}~\bibnamefont{Lange}},
  \bibinfo{author}{\bibfnamefont{A.}~\bibnamefont{Volodin}},
  \bibinfo{author}{\bibfnamefont{R.}~\bibnamefont{Szymczak}}, \bibnamefont{and}
  \bibinfo{author}{\bibfnamefont{V.}~\bibnamefont{Moshchalkov}},
  \bibinfo{journal}{Nature Mater.} \textbf{\bibinfo{volume}{3}},
  \bibinfo{pages}{793} (\bibinfo{year}{2004}).

\bibitem[{\citenamefont{Tamura and Endo}(1969)}]{Tamura1969}
\bibinfo{author}{\bibfnamefont{K.}~\bibnamefont{Tamura}} \bibnamefont{and}
  \bibinfo{author}{\bibfnamefont{H.}~\bibnamefont{Endo}},
  \bibinfo{journal}{Phys. Lett. A} \textbf{\bibinfo{volume}{29}},
  \bibinfo{pages}{52} (\bibinfo{year}{1969}).

\bibitem[{\citenamefont{Blatter et~al.}(1994)\citenamefont{Blatter, Feigel'man,
  Geshkenbein, Larkin, and Vinokur}}]{Blatter94}
\bibinfo{author}{\bibfnamefont{G.}~\bibnamefont{Blatter}},
  \bibinfo{author}{\bibfnamefont{M.~V.} \bibnamefont{Feigel'man}},
  \bibinfo{author}{\bibfnamefont{V.~B.} \bibnamefont{Geshkenbein}},
  \bibinfo{author}{\bibfnamefont{A.~I.} \bibnamefont{Larkin}},
  \bibnamefont{and} \bibinfo{author}{\bibfnamefont{V.~M.}
  \bibnamefont{Vinokur}}, \bibinfo{journal}{Rev. Mod. Phys.}
  \textbf{\bibinfo{volume}{66}}, \bibinfo{pages}{1125} (\bibinfo{year}{1994}).

\end{thebibliography}

\end{document}